\documentclass{iau} 
\usepackage{amsmath,amssymb,graphicx}

\title[Stellar population synthesis with CEM] 
{Stellar population synthesis of galaxies with chemical evolution model}

\author[Shiyin Shen \& Jun Yin]   
{Shiyin Shen$^{1,2}$  \and Jun Yin$^1$  }

\affiliation{$^1$ Key Laboratory for Research in Galaxies and Cosmology, Shanghai
Astronomical Observatory, Chinese Academy of Sciences, 80 Nandan Road, Shanghai,
200030, China\\
 $^2$Key Lab for Astrophysics, Shanghai, 200234, China;\\
email: {\tt ssy@shao.ac.cn.cn}}

\pubyear{2019}
\volume{341}  
\jname{Challenges in Panchromatic Galaxy Modelling with Next Generation Facilities}
\editors{ M. Boquien, E. Lusso, C. Gruppioni, and P. Tissera.}
\begin{document}

\maketitle

\begin{abstract}
The derivation of accurate stellar populations of galaxies is a non-trivial task because of the well-known age-metallicity degeneracy. We aim to break this degeneracy by invoking a chemical evolution model(CEM) for isolated disk galaxy, where its metallicity enrichment history(MEH) is modelled to be tightly linked to its star formation history(SFH). Our CEM has been successfully tested on  several local group dwarf galaxies whose  SFHs and MEHs have been both independently measured from deep color-magnitude diagrams of individual stars. By introducing the CEM into the stellar population fitting algorithm as a prior, we expect that the SFH  of galaxies could be better constrained.
\end{abstract}

\firstsection 
\section{Introduction}

Star formation history(SFH), i.e. the amount of stars formed in galaxies as function of time, is one of the elements that describes galaxy evolution. Stars are formed from  cool phase gas and then evolved stars return metals into the interstellar medium(ISM).  ISM cools and new generation of stars formed. Among this circle, SFH is the key factor that links the gas cooling and metallicity enrichment history(MEH) of galaxies.

However, recovering SFH of galaxies is difficult(\cite{Conroy2013}). For very nearby galaxy, the deep color-magnitude diagram(CMD) reaching the  turn-off stars is known as the only direct and most reliable method that can recover both of its detailed SFH and MEH. For galaxy at larger distance with only  integrated stellar light observed, stellar population synthesis methods have been developed and used to fit its spectral energy distribution(SED) . In an idealized case, the SED of a galaxy can be viewed as a composition  of  single stellar populations(SSPs) with different ages and metallicities. However, because of the very similarities of the old SSPs($t > 1$ Gyr) and  the age-metallicity degeneracy among SSPs, the recovering of the detailed weights of each SSPs, especially for those with age older than 1 Gyr, is very difficult. In reality, the observables are further complicated by many other details, e.g. stellar kinematics, ionized gas emissions, central AGNs, dust attenuation and emission etc. Therefore, typically,  only the first order description of the SFH, e.g. the average age or the fraction of young stellar population($t <1$ Gyr), rather than its detailed shape,  could be reliably derived from the observed SEDs of galaxies.

\section{Recovering detailed SFH}

In popular SED fitting algorithms(e.g. STARLIGHT, \cite{Cid05}), the ages and metallicties of SSPs are considered as independent parameters. To recover  a detailed  SFH and MEH of a galaxy, a library of SSPs with different ages and metallicities is first built,  then the best combinations of these SSPs are searched in the huge library space. However, because of the age-metallicity degeneracy,  the recovering of the SSP weights from SED is an ill-posed problem. 

To show this problem more intuitively, we make a mock galaxy spectrum and test how good a full spectrum fitting can recover its SFH and MEH. We take the SFH and MEH of the LMC bar region, which are derived from deep CMD fitting(\cite{LMC}) and are shown as the solid circles in Fig. 1, to build the mock spectrum.  We use SSPs from MILES library (\cite{miles}) to generate the  mock spectrum in the wavelength range $3600-7500\AA$ with  resolution  $R=2000$ and S/N=50 per pixel. For simplicity, we neither consider the kinematics of stellar populations nor reshape the spectrum with dust attenuation.

For this idealized case, whether can we recover the input SFH and MEH accurately using a full spectrum fitting? We take the same MILES SSP library and use a Monte Carlo Markov Chain to probe the full age and metallicity space  and search for the best solution. The resulted SFH and MEH, converted from the likelihood distributions of SSP weights,  are shown as the red squares in each panel of Fig. 1. As can be seen, the full spectrum fitting algorithm  can not give a good constraint on either MEH or SFH. 

Is an accurate MEH knowledge helpful to recover the accurate SFH? To test this idea, we set the metallicities of the different age SSPs to follow the input MEH(the blue circles in the bottom panel of Fig. 1). We then make the full spectrum fitting again. Now, we only need to probe the weights of the SSPs in age space. The resulted SFH is shown as the green crosses in the top panel of Fig. 1. As can be seen, when MEH is known as a prior, the SFH is recovered with much better accuracy. 

Fig. 1 shows  that the prior information on the MEH, which breaks  the age-metallicity degeneracy, is a key factor to recover the accurate SFH of galaxies.
Actually, the importance of the prior information on SFH or MEH have already been explored by many SED fitting codes. For example, the code STECKMAP(\cite{steckmap}) allows a penalization of the best fitting with an assumed MEH. However, a realistic MEH needs physical justification. That is the chemical evolution model we will discuss next.

\begin{figure}
\centering
\includegraphics[width=120mm]{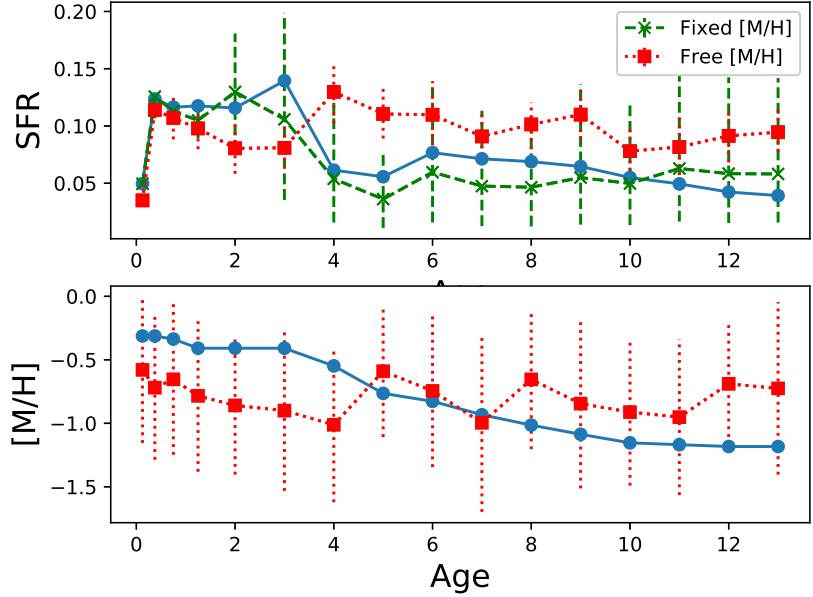}
\caption{SFH(top panel) and MEH(bottom panel) of a mock galaxy. The solid circles represent the input SFH and MEH, whereas the red squares show the SFH and MEH recovered from the mock spectrum with no prior information. The green crosses in the top panel show the recovered SFH when its MEH is assumed to be the blue circles in the bottom panel.}
\end{figure}

\section{Chemical evolution model}

We consider  galaxy as a  pool of stars and gas. At any given time, there are both inflow and outflow of gas into the pool. The inflow gas is primordial, i.e. with zero metallicity. Stars formed from gas, and died stars return enriched gas into the pool. Some of the enriched gas makes up of the outflows.

We write the time dependent star formation rate(SFR), gas inflow and outflow rate  as $\Psi(t)$, $f_{in}(t)$ and $f_{out}(t)$ respectively. Stars are born from gas pool following  the initial mass function(IMF).We assume that the massive stars($M > 1 M_\odot$) die immediately  and return gas into the pool, while the low mass stars($M < 1 M_\odot$) have infinite lifetime. For the classical Salpter IMF, the gas return fraction $R$ is $\sim0.3$. Thus, the change rate of gas in pool is 

\begin{equation}
    \frac{d M_{gas}(t)}{dt} = -(1-R)\Psi(t) + f_{in}(t) - f_{out}(t)\,.
\end{equation}
On the other hand, the surface star formation rate density of a galaxy is known to be tightly correlated with its surface gas density $\Sigma_{gas}$, i.e. the well-known Kennicutt-Schmidt relation, $\Psi_\Sigma \propto \Sigma_{gas}^{1.4}$(\cite{Kennicutt98}). For gas outflow, we assume it is driven by supernova explosion so that it is proportional to SFR and inversely correlated with the potential well of a galaxy,
\begin{equation}\label{eq:fout}
    f_{out}(t)=\eta \left[0.5+\left(\frac{v_{vir}}{70 ~km~s^{-1}}\right)^{-3}\right]\cdot\Psi(t) 
\end{equation}
where $v_{vir}$ is the circular velocity of galaxy halo, and $\eta$ is the wind efficiency.

The evolution of metallicity $Z(t)$ is a balance between the star formation and metal outflow, which is written as
\begin{equation}\label{eq:z}
  \begin{split}
    \frac{d(Z M_{gas})}{dt}&=-Z(1-R)\Psi(t)+y(1-R)\Psi(t)-Zf_{out} \,,
  \end{split}
\end{equation}
 where $y$ is the yield and we take  $y=0.1$.

With above equations (3.1 to 3.3), once with structure parameters(to convert gas mass to surface density), for any given $\Psi(t)$ of a galaxy, we can predict both its $Z(t)$ and  gas inflow/outflow histories.

\section{Test on LCID galaxies}

We test above CEM with LCID galaxies(\cite{LCID}), whose detailed SFHs and MEHs have been obtained  using deep CMDs from HST.  As an example, we take the SFHs of three different type LCID galaxies, Tucana(dSph), LGS-3(dTran), IC 1613(dIrr), and predict their $Z(t)$ from the above CEM. We calculate their average surface densities inside the half-light radii and use their stellar masses to estimate the circular velocities. The wind efficacy is set to be $\eta=0.4$ in all the cases. The results are shown in Fig. 2. As can be seen, for all three different galaxies, our CEM reproduces their MEHs  from SFHs quite well.

\begin{figure}
\centering
\includegraphics[width=120mm]{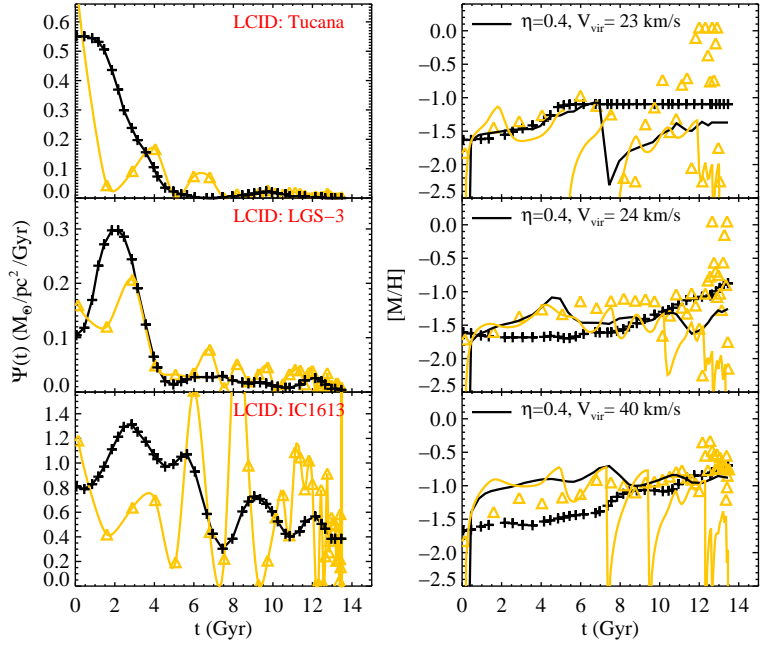}
\caption{SFH(left panels) and MEH(right panels) of 3 LCID galaxies. The black crosses  show the best estimations of SFHs and MEHs from LCID project,whereas the yellow triangles show one of their Monte-Carlo realization(indicating the uncertainty, \textit{Dan Weisz, private communication.}). The solid lines in the left panels are the continuous SFHs used as the input of CEM, while the model predicted MEHs are shown as the  solid curves with corresponding colors in right panels.  }
\end{figure}

\section{Conclusion}
Encouraged by Fig. 2, we believe that our CEM can be used to break the age-metallicity degeneracy in stellar population synthesis studies.  Specifically, we may start from the SED fitting with both SFH and MEH free. New MEH then is predicted from preliminary SFH using CEM.  With several iterations, a self-consistent SFH and MEH would be finally obtained. We expect that this algorithm would reduce the uncertainties of the final SFH estimation, and is proper for isolated disk galaxies.

\end{document}